\documentclass[sigconf, screen]{acmart}

\usepackage{amsmath,amsfonts,bm}

\def\eqref#1{equation~\ref{#1}}

\def\1{\bm{1}}

\DeclareMathAlphabet{\mathsfit}{\encodingdefault}{\sfdefault}{m}{sl}
\SetMathAlphabet{\mathsfit}{bold}{\encodingdefault}{\sfdefault}{bx}{n}

\newcommand{\bx}{\mathbf{x}}
\newcommand{\by}{\mathbf{y}}

\usepackage{multirow}

\usepackage{enumitem}
\newcommand{\methodname}{{Super Star}}

\AtBeginDocument{%
  }

\setcopyright{acmlicensed}
\copyrightyear{2018}
\acmYear{2018}
\acmDOI{XXXXXXX.XXXXXXX}
\acmConference[Conference acronym 'XX]{Make sure to enter the correct
  conference title from your rights confirmation email}{June 03--05,
  2018}{Woodstock, NY}
\acmISBN{978-1-4503-XXXX-X/2018/06}

\begin{document}

\title{Super Star: Towards Streaming Real-time Interactive Agents for Digital Humans}

\author{Wentao Jiang}
\affiliation{%
  \institution{ShanghaiTech University}
  \city{Shanghai}
  \country{China}}
\email{jiangwt2024@shanghaitech.edu.cn}

\author{Youchen Xie}
\affiliation{%
  \institution{LIGHTSPEED}
  \city{Shenzhen}
  \country{China}
}
\email{youchenxie@tencent.com}

\author{Haidi Fan}
\affiliation{%
  \institution{LIGHTSPEED}
  \city{Shenzhen}
  \country{China}
}
\email{haidifan@tencent.com}

\author{Yajing Chen}
\affiliation{%
  \institution{LIGHTSPEED}
  \city{Shenzhen}
  \country{China}
}
\email{jadeyjchen@tencent.com}

\author{Xin Wang}
\affiliation{%
  \institution{LIGHTSPEED}
  \city{Shenzhen}
  \country{China}
}
\email{alexinwang@tencent.com}

\author{Ye Shi}
\affiliation{%
  \institution{ShanghaiTech University}
  \city{Shanghai}
  \country{China}}
\email{shiye@shanghaitech.edu.cn}

\author{Jingya Wang}
\authornote{ Corresponding author. \\
This work was supported by HPC Platform of ShanghaiTech University.
}
\affiliation{%
  \institution{ShanghaiTech University}
  \city{Shanghai}
  \country{China}}
\email{wangjingya@shanghaitech.edu.cn}

\renewcommand{\shortauthors}{Trovato et al.}

\begin{abstract}
Existing co-speech gesture generation methods are predominantly studied in offline settings, where gestures are synthesized from complete speech segments. However, interactive digital humans in real-world scenarios are required to generate speech-synchronous gestures online, using only currently available response audio under strict latency constraints. As a result, prior methods are unsuitable for real-time interaction, as they either rely on future speech information or incur substantial inference delay. In this paper, we formulate online co-speech gesture generation for interactive digital humans and propose a real-time interactive framework that couples a streaming speech response module with an online gesture generation module. Specifically, the gesture generator is designed as a causal multimodal autoregressive model that predicts body motion from streaming response speech and motion history, enabling low-latency and speech-aligned gesture synthesis without access to future speech. To support this setting, we further propose an offline data synthesis pipeline tailored to virtual companion scenarios, which leverages topic- and emotion-aware subject corpora to construct diverse human-agent dialogues and then generates co-speech gestures conditioned on the agent responses. Moreover, to bridge the gap between offline data construction and online deployment, we establish a self-evolving training loop by incorporating user feedback collected during online interaction into the data generation process, enabling continual adaptation to user preferences. Extensive experiments demonstrate that our framework achieves superior better latency-quality trade-off, stronger speech-motion synchronization, and higher user preference than competitive existing baselines. Project Page: \url{https://super-star-2026.github.io/}
\end{abstract}

\ccsdesc[300]{Computing methodologies~Motion processing}
\ccsdesc[300]{Computing methodologies~Activity recognition and understanding}

\keywords{Online Co-speech Gesture Generation, Digital Humans, Interactive Agents, Self-evolution}
\begin{teaserfigure}
  \centering
  \includegraphics[width=0.82\textwidth]{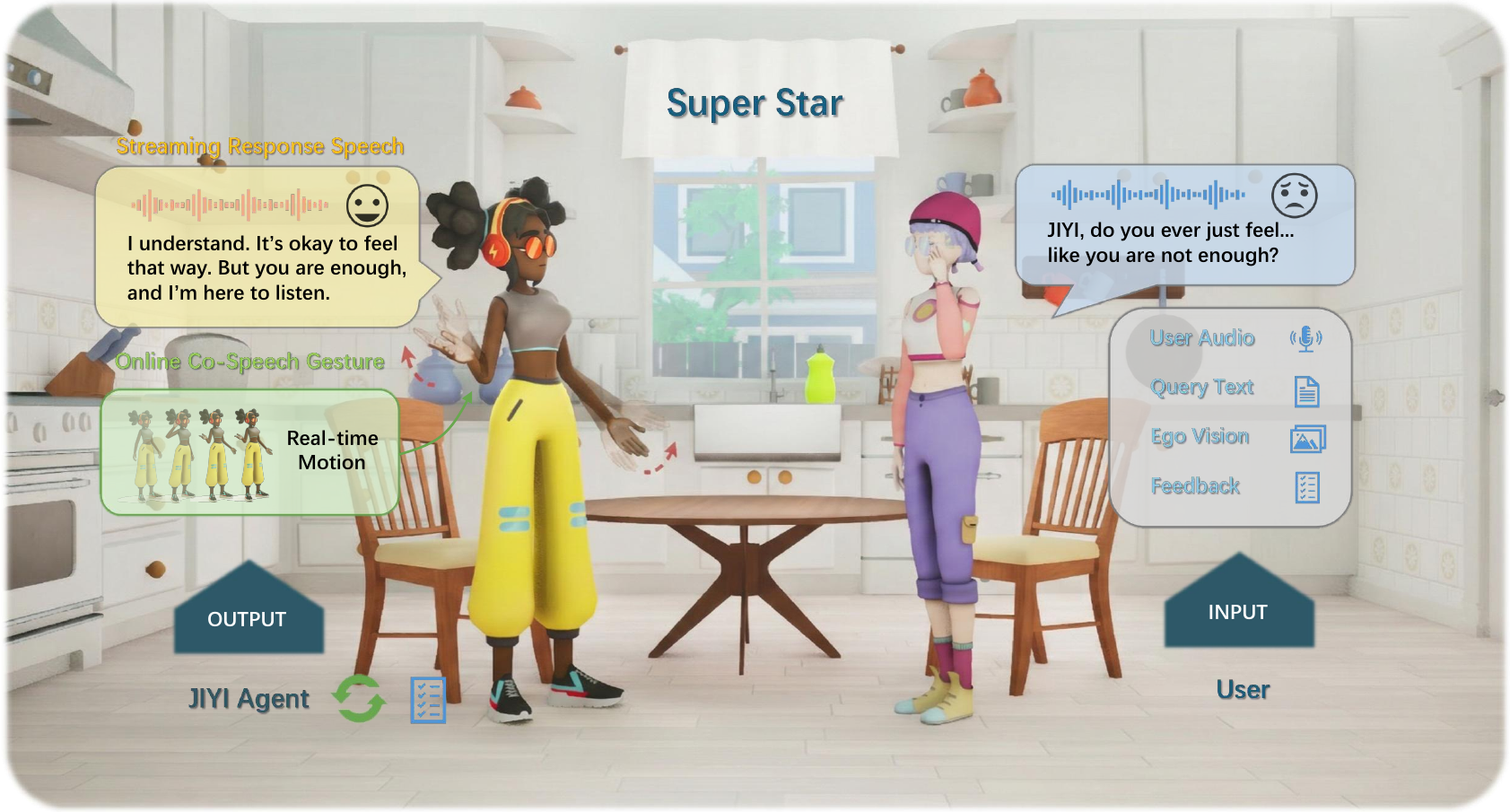}
  \caption{\methodname~enables 3D digital humans to interact with users in real-time and generate speech synchronization gestures online based on user multimodal input via an online real-time interactive pipeline, which is trained through our closed-loop self-evolving data pipeline to support continual adaptation to user preferences.}
  \Description{Enjoying the baseball game from the third-base
  seats. Ichiro Suzuki preparing to bat.}
  \label{fig:teaser}
\end{teaserfigure}


\maketitle
\section{Introduction}
\label{sec:intro}
Co-speech gesture generation~\cite{nyatsanga2023comprehensive, liu2022beat, ghorbani2023zeroeggs} aims to synthesize natural body motion that are temporally aligned with speech, and has been widely studied for digital humans, embodied conversational agents, and virtual avatars~\cite{zhang2025vibes,huang2025live,jiang2025solami,wampfler2025platform,cai2025digital}. Recent advances have achieved promising results in modeling speech-motion correspondence. However, most existing methods~\cite{liu2024emage, chen2025language,chen2024enabling} are designed for \textbf{offline settings}, where the model has access to the complete speech segment before gesture generation. While effective for pre-recorded animation, this assumption limits their applicability in real-world interaction scenarios. In practice, digital humans are required to respond to users in real time and produce speech-synchronous gestures online, using only currently available response audio. This motivates the problem of \textbf{online co-speech gesture generation}, a more challenging setting where future speech is unavailable and low latency is essential.

Directly applying existing offline methods to this online setting is often ineffective. On the one hand, many prior approaches~\cite{liu2024emage, chen2025language,chen2024enabling} rely on full-sequence speech context or future acoustic cues to generate temporally coherent and semantically appropriate gestures. On the other hand, methods that process long speech segments typically introduce substantial inference delay, which breaks immediacy required in live interaction. Beyond the modeling challenge, another practical bottleneck lies in data availability: existing co-speech motion datasets are largely collected from monologue~\cite{liu2022beat,ghorbani2023zeroeggs} or general social scenarios~\cite{ng2024audio,mughal2024convofusion,mclean2025embody,qi2025co,zhang2025vibes} and provide limited support for \textbf{interactive virtual companion} settings, where responses are highly diverse in topic and emotion, and gestures require richer posture variation and stronger affective expression to provide emotional value and companionship. To address these issues, we propose a real-time interactive framework for online co-speech gesture generation and a close-loop pipeline that combines offline interactive data synthesis with a user-feedback-driven self-evolution loop to support training and continual improvement in this setting.

Specifically, our framework consists of two tightly coupled components: a streaming speech response module and an online gesture generation module. Given multimodal user inputs, the response module first produces the agent response speech in a streaming manner, providing immediately playable audio for real-time interaction. Conditioned on the streaming response speech and previously generated motion history, the online gesture generator then predicts body motion causally in an autoregressive manner, so that the digital human can output speech-synchronous gestures without waiting for the complete utterance. To support training in virtual companion scenarios, we further construct an offline interactive data synthesis pipeline that generates diverse human-agent dialogues and corresponding co-speech motions, and we connect this pipeline with online user feedback to form a self-evolving loop for continual adaptation to user preferences.

An overview of the proposed framework is shown in Figure~\ref{fig:pipeline1}. During online interaction, the system first receives user inputs and generates streaming response speech through the response module. The online gesture generator then takes the streaming speech tokens together with motion history as input and predicts the current body motion in a strictly causal manner. During offline training, we build an interactive data synthesis pipeline tailored to virtual companion scenarios, where topic- and emotion-aware subject corpora are used to construct diverse dialogues and synthesize paired co-speech motion data. The collected online interaction data and user feedback are further fed back into this pipeline to refine the training set and continuously improve the deployed model. Our main contributions are as follows: 
\vspace{-0.8mm}
\begin{itemize}[leftmargin=*]
    \item We formulate online co-speech gesture generation for interactive digital humans, a practically important setting that requires causal gesture generation from streaming response speech without future speech access and under real-time latency constraints.
    \item We propose a causal online gesture generator tailored to this setting. Specifically, we design a causal audio-conditioned cross-attention mechanism to enforce online speech-motion alignment during training, and a cross-modal autoregressive factorization over body parts without future context.
    \item We propose a closed-loop self-evolving data pipeline tailored to virtual companions, 
    where preferred online interaction samples are reused as supervision and in-context guidance for the next iteration, enabling continual adaptation to user preferences.
    \item Extensive experiments under strict online protocols demonstrate that the proposed framework achieves superior latency-quality tradeoff, speech-motion synchronization, and human preference over existing baselines.
\end{itemize}
\section{Related Work}
\label{sec:related}
\subsection{Offline Co-Speech Gesture Generation}
Co-speech gesture generation~\cite{nyatsanga2023comprehensive, liu2022beat, ghorbani2023zeroeggs,lee2019talking,zhou2022responsive} aims to synthesize natural body movements that are temporally aligned with speech, and has been extensively studied for digital humans, embodied conversational agents and virtual avatars. Early works mainly focused on generating upper-body~\cite{liu2022disco,yi2023generating} or pose-level~\cite{ginosar2019learning,shlizerman2018audio} gestures from speech audio~\cite{habibie2021learning}, text~\cite{guo2022generating}, or their combination, using recurrent networks~\cite{shlizerman2018audio}, variational models~\cite{li2021audio2gestures}, or adversarial training~\cite{liu2022disco,liu2022learning} to capture speech-motion correspondence~\cite{ahuja2020style,yi2023generating,yoon2020speech, alexanderson2020style} including semantics~\cite{zhang2024semantic,liu2025semges, zhang2025semtalk}. With the development of large-scale multimodal datasets and stronger generative models~\cite{xu2024mambatalk,cheng2025hop,wang2026mmofusion}, recent approaches have achieved substantial progress in generating more realistic, diverse, facial~\cite{fan2022faceformer,xing2023codetalker,chu2025unils,yang2026streamingtalker,ng2022learning} and expressive full-body co-speech motions~\cite{zhang2024semantic, liu2024emage, chen2025language,chen2024enabling,liu2025gesturelsm}. These methods typically leverage transformer-based architectures, diffusion models~\cite{cheng2025holegest,cheng2024siggesture,ao2023gesturediffuclip,yang2023diffusestylegesture,zhu2023taming,chhatre2024emotional, yang2023diffusestylegesture}, or discrete motion tokenization~\cite{guo2024momask,zhang2023generating} to model the mapping from speech to gesture.

Despite their strong performance, most existing methods are designed for \emph{offline generation}, where the complete speech segment is available before motion synthesis. Such a formulation is well suited to pre-recorded animation or post-processing scenarios, but is less compatible with interactive digital humans, where gesture generation is required to be performed causally from streaming response speech under strict real-time latency constraints. In contrast to these prior works, we explicitly study \emph{online co-speech gesture generation}, where future speech is unavailable and low-latency generation is essential.

\subsection{Multimodal Interactive Systems for 3D Digital Humans}

Recent multimodal interactive systems for 3D digital humans aim to build socially intelligent embodied agents through multimodal perception~\cite{hurst2024gpt}, reasoning~\cite{jiang2023motiongpt}, and response generation~\cite{jiang2025solami,zhang2025vibes,cai2025towards, kim2024body}. Related works on reaction synthesis~\cite{xu2024regennet,jiang2025arflow}, social modeling~\cite{zhang2025social,li2026interagent}, and streaming motion generation further move toward interactive and deployable 3D agents~\cite{liu2024physreaction,xu2024regennet,jiang2025arflow,cai2025flooddiffusion,ji2025towards}. To the best of our knowledge, we note that some recent studies have started to explore real-time~\cite{zhan2026umo,deng2026u}, streaming~\cite{xiao2025motionstreamer}, or deployable embodied motion generation~\cite{zhang2026proact,ji2026discoforcing} more explicitly. However, these works are generally not directly comparable to our setting, since they typically focus on broader motion generation rather than full-body strictly online co-speech gesture generation conditioned on streaming response speech, and therefore cannot directly satisfy the low-latency speech-gesture synchronization requirement in real-time digital humans. 

\subparagraph{\textbf{Datasets.}}
Meanwhile, existing conversational and interactive behavior datasets, such as audio-driven conversational embodiment datasets~\cite{ng2024audio,mughal2024convofusion} and broader multimodal interaction corpora~\cite{jiang2025solami,mclean2025embody,qi2025co,zhang2025vibes}, are mostly collected from social communication or general interaction scenarios~\cite{intergen,xu2023inter,zhang2024hoi}. Such datasets are still insufficient for our virtual companion scenarios for two reasons: first, the synchronization between response speech and body motion is often not designed for strict online co-speech generation; second, virtual companion scenarios require richer posture variation and stronger affective expression to provide emotional value and companionship, which are not fully covered by existing social-scene data. Motivated by these limitations, we introduce an offline interactive data synthesis pipeline and employ motion capture systems to collect high-quality co-speech motion sequences captured from professional actors we invite.

\section{Method}
\label{sec:flow}

\begin{figure*}[t]
  \centering 
  \includegraphics[width=0.99\linewidth]{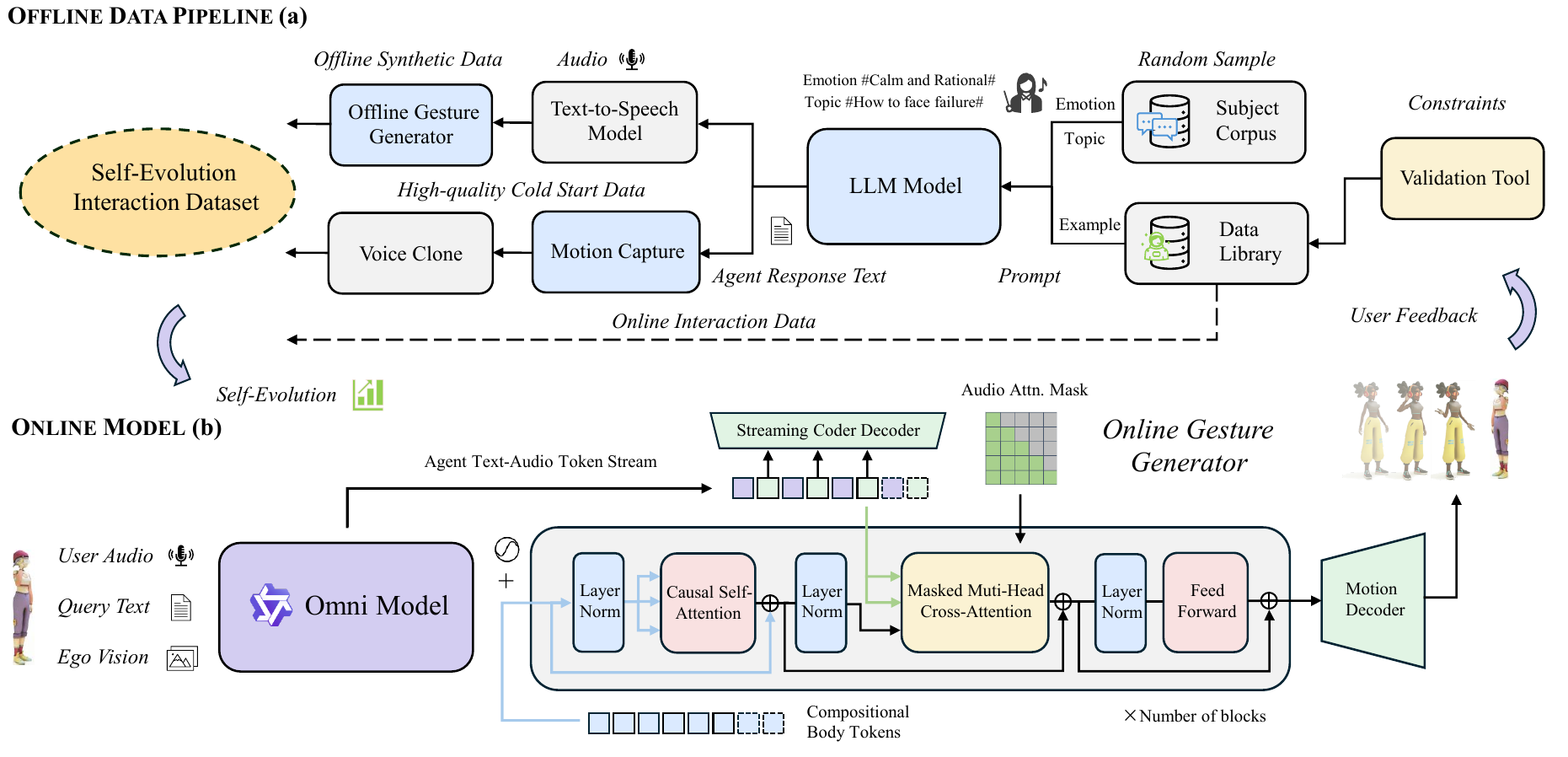}
  \caption{Overview of \methodname. (a) Offline interactive data synthesis and self-evolving loop pipeline for constructing virtual-companion-oriented training data that continuously adapts to user preferences. (b) Online real-time interaction pipeline, where a streaming speech response module is coupled with our online gesture generator to produce low-latency co-speech motions.}
  \label{fig:pipeline1}
\end{figure*}

\subsection{System Overview}
\label{sec:overview}

As shown in Figure~\ref{fig:pipeline1}, we propose a unified framework for streaming real-time interactive digital humans. The framework consists of two interconnected parts: an \textbf{online real-time interaction pipeline} and an \textbf{offline interactive data synthesis pipeline}. The online pipeline enables the digital human to generate speech-synchronous gestures causally from streaming response speech, while the offline pipeline provides interaction-oriented training data for virtual companion scenarios and supports continual model improvement through a self-evolving loop.

In the online pipeline, a response module first produces streaming agent speech from multimodal user input. Conditioned on the streaming response speech and previously generated motion history, an online gesture generation module then predicts body motion incrementally under strictly causal constraint, enabling low-latency and synchronized gestures without future speech access.

In the offline pipeline, we synthesize virtual-companion-oriented training data by constructing diverse human-agent dialogues from topic- and emotion-aware subject corpora and generating corresponding co-speech motions conditioned on the agent responses. Online interaction data and user feedback are further fed back into this pipeline to refine the training set and continuously adapt the deployed model.

We describe the main components of the framework in the following sections: Section~3.2 introduces the streaming speech response module, Section~3.3 presents the online gesture generation model, and Section~3.4 details the offline data synthesis and self-evolution pipeline.

\subsection{Streaming Speech Response Generation}
\label{sec:speech}
The omni-modal model processes user audio, text or visual information to generate audio responses. We use the Qwen-Omni streaming real-time model~\cite{xu2025qwen2,xu2025qwen3}, which can output real-time response speech in a streaming manner. After the user inputs multimodal information, the model generates token-level agent audio in a streaming fashion for immediate playback. For our lower system latency, at the same time that these streaming audio tokens are decoded into speech by the streaming codec decoder, we use them as our conditional input for our subsequent online co-speech gesture generation module.

\subsection{Online Co-Speech Gesture Generation}
\label{sec:co-speech}

In real-time interaction, digital humans are required to generate co-speech gestures \emph{online}, i.e., to predict the current body motion causally from the streaming response speech without access to future audio. Formally, given the speech sequence observed up to the current frame $H$, denoted as $\by=\{y^i\}_{i=1}^{H}$, the goal is to generate corresponding co-speech motion $\bx=\{x^i\}_{i=1}^{H}$ at low latency while preserving motion naturalness and speech-motion synchronization. This setting is more challenging than conventional offline co-speech generation, since the model must make temporally coherent motion predictions under incomplete speech context. To address this problem, we propose a causal multimodal autoregressive gesture generator built upon a compositional motion tokenization.
\subparagraph{\textbf{Compositional body motion tokenization.}}
A key challenge of online gesture generation lies in the ambiguity of motion prediction when only partial speech observations are available. Directly modeling holistic full-body motion in this setting is difficult, as different body parts exhibit distinct temporal dynamics and respond differently to speech cues. To make online prediction more tractable while preserving full-body coordination, we adopt a compositional motion representation following~\cite{liu2024emage,chen2025language}. 

Specifically, we divide the body into semantically meaningful parts represented by 6D rotations, including lower body $\mathbf{g}_l$, upper body $\mathbf{g}_u$, hands $\mathbf{g}_h$ and head-related facial expressions $\mathbf{g}_f$. The motion sequence is thus represented as
$
\bx=\{\mathbf{g}_f,\mathbf{g}_h,\mathbf{g}_u,\mathbf{g}_l\}.
$

Based on this compositional representation, we train four separate VQ-VAEs~\cite{van2017neural,zhang2023generating} to tokenize each body part independently. For each part, the VQ-VAE encoder $\mathcal{E}$ adopts a four-layer temporal convolutional network (TCN) to extract continuous latent motion features $\mathbf{z}^{1:T}=\mathcal{E}(\mathbf{g}^{1:T})$, which are then quantized into discrete motion tokens via
\begin{equation}
\mathbf{q}^t = Q(\mathbf{z}^t) := \mathop{\arg\min}_{\mathbf{q}^k \in Q} \|\mathbf{z}^t - \mathbf{q}^k\|^2 ,
\end{equation}
where $\mathbf{q}^t$ denotes the discrete code assigned to the latent feature $\mathbf{z}^t$ and 
$
Q = \{\mathbf{q}_f, \mathbf{q}_h, \mathbf{q}_u, \mathbf{q}_l\}
$ is the quantized motion latent space.
The decoder $\mathcal{D}$ reconstructs the motion from the quantized tokens as $\hat{\mathbf{g}}^{1:T}=\mathcal{D}(\mathbf{q}^{1:T})$.
Following~\cite{chen2025language}, we train the tokenizer using reconstruction, temporal smoothness, mesh, and commitment losses:
\begin{equation}
\begin{aligned}
\mathcal{L}_{\text{total}} = &\mathcal{L}_{\text{rec}}(\mathbf{g}, \hat{\mathbf{g}}) + \mathcal{L}_{\text{vel}}(\mathbf{g}', \hat{\mathbf{g}}') + \mathcal{L}_{\text{acc}}(\mathbf{g}'', \hat{\mathbf{g}}'') + \mathcal{L}_{\text{mrec}}(\mathbf{g}, \hat{\mathbf{g}}) \\
&+ \mathcal{L}_{\text{mvel}}(\mathbf{g}', \hat{\mathbf{g}}') + \mathcal{L}_{\text{macc}}(\mathbf{g}'', \hat{\mathbf{g}}'') + \mathcal{L}_{\text{comm}}(\mathbf{g}, \mathbf{q})~,
\end{aligned}
\end{equation}
where $\hat{\mathbf{g}}'$ and $\hat{\mathbf{g}}''$ denote the first-order and second-order temporal derivatives of the reconstructed motion $\hat{\mathbf{g}}$, respectively. 
Following~\cite{chen2025language}, we use Geodesic loss~\cite{tykkala2011direct} for the pose reconstruction loss $\mathcal{L}_{\text{rec}}$, $l_1$ loss for the velocity/acceleration losses of pose and mesh ($\mathcal{L}_{\text{vel}}$, $\mathcal{L}_{\text{acc}}$, $\mathcal{L}_{\text{mvel}}$, and $\mathcal{L}_{\text{macc}}$), and $l_2$ loss for the mesh reconstruction loss $\mathcal{L}_{\text{mrec}}$ and the codebook commitment loss $\mathcal{L}_{\text{comm}}$. 

This compositional tokenization is particularly suitable for online generation. By decomposing motion into multiple semantically meaningful parts, it reduces the complexity of causal prediction in an incomplete speech context and allows the generator to model heterogeneous motion patterns while maintaining global body consistency.

\subparagraph{\textbf{Online multimodal autoregressive modeling.}}
Most existing co-speech gesture generation methods~\cite{liu2024emage,chen2024enabling,chen2025language} are designed for offline generation and assume access to complete speech segments. Such a formulation is not applicable to real-time interaction, where future speech is unavailable and waiting for the complete response segment would introduce unacceptable latency. We therefore formulate online co-speech gesture generation as a causal autoregressive prediction problem over the quantized motion tokens.

At each time step, the model predicts the current motion token conditioned on the available response speech stream and the previously generated motion tokens. Since different body parts are tokenized independently, naively predicting them separately may lead to inconsistent full-body motions. To preserve holistic coordination, we explicitly model the dependencies among compositional motions through cross modal autoregressive generation. Let $\mathbf{C}_{1:\tau}:=(C_{1:\tau}^f, C_{1:\tau}^h, C_{1:\tau}^u, C_{1:\tau}^l)$ denote the token sequences of all body parts up to time $\tau$. We model their joint distribution conditioned on the observed speech features $Y_{1:\tau}$ as
\begin{equation}
p(\mathbf{C}_{1:\tau} \mid Y_{1:\tau})
=
\prod_{t=1}^{\tau}
\prod_{m \in \{f,h,u,l\}}
p(c_t^m \mid \mathbf{c}_{<t}, \mathbf{y}_{\le t}),
\end{equation}
where $\mathbf{c}_{<t}:=(\mathbf{c}_{<t}^f,\mathbf{c}_{<t}^h,\mathbf{c}_{<t}^u,\mathbf{c}_{<t}^l)$ represents the motion history from all body parts before time $t$, and $\mathbf{y}_{\le t}$ denotes the speech observations available up to the current time step. This factorization enables the model to exploit temporal dependency from past motions while leveraging the mutual information across body parts. Such cross modal compositional modeling is especially important in the real-time online setting, where each body part can be predicted in parallel and future speech is not available to stabilize full-body motion prediction.

\subparagraph{\textbf{Causal masked audio-conditioned cross-attention.}}
To align body motion generation with streaming speech, we adopt an audio-conditioned cross-attention mechanism between the speech token stream and the motion token stream, as shown in Figure~\ref{fig:pipeline1}. Unlike offline cross-attention, however, online generation requires strict causality: each motion prediction should only depend on the current and past speech observations. To satisfy this requirement, we introduce a causal mask into the multi-head cross-attention layer, so that the motion token at time $t$ can only attend to speech features in $\mathbf{y}_{\le t}$ and is prevented from accessing any future speech information.

This design is not merely an inference-time masking strategy. We apply the same causal constraint during training, forcing the model to learn gesture dynamics directly from streaming speech rather than relying on unavailable future cues. As a result, the proposed cross-attention mechanism serves as an online cross-modal alignment module that improves speech-motion synchronization while remaining fully compatible with real-time deployment. In practice, it also allows the model to react immediately to newly arrived speech tokens, leading to a better latency-quality trade-off than existing baselines.

During training, the online gesture generator is optimized with a token prediction objective over the ground-truth motion token sequence under the causal constraint, and the predicted tokens are then decoded into continuous motion through pretrained VQ-VAE decoders. 
During inference, the model incrementally generates motion tokens from streaming response speech and motion history, enabling low-latency co-speech gestures in real-time interaction.

\subsection{Offline Interactive Data Synthesis}
\label{sec:offline}
Although the online gesture generator is designed for real-time deployment, its performance critically depends on the availability of interaction-oriented training data. However, existing co-speech motion datasets are mostly collected from monologue or general social scenarios, and therefore provide limited support for interactive virtual companion settings. In such scenarios, agent responses are often highly diverse in topic and emotion, and the associated co-speech gestures require to reflect not only speech rhythm but also communicative intent and affective state. To bridge this gap, we introduce an offline interactive data synthesis pipeline that constructs virtual-companion-oriented training data for online co-speech gesture generation, and further connects offline data construction with online deployment through a self-evolving loop in Figure~\ref{fig:pipeline1}.

\subparagraph{\textbf{Subject-corpus-guided dialogue synthesis.}}
A direct way to construct interactive training data is to prompt a large language model to generate user-agent dialogues. However, naive prompting often leads to limited diversity and repetitive interaction patterns, due to the inherent inductive bias of LLMs. Especially in virtual companion scenarios, user-agent dialogues should cover a broad range of topics and emotional states. To improve controllability and diversity, we build a \emph{subject corpus}
$
\mathcal{S}=\{t_0,t_1,\ldots,t_m,e_0,e_1,\ldots,e_n\},
$
which consists of domain-specific text segments describing dialogue subjects, including topic and emotion cues.

During synthesis, we randomly sample a topic segment $t_i$ and an emotion segment $e_j$ from $\mathcal{S}$ and use it as structured context to prompt a powerful LLM to generate a human-agent dialogue. For example, given a topic segment such as \emph{``how to face failure''} and an emotion segment such as \emph{``calm and rational''}, the LLM is prompted to generate a related dialogue where the agent is defined as a life mentor to comfort and provide rational advice to the user in a calm tone. This subject-corpus-guided strategy improves the coverage and diversity of synthesized interactions and better matches the requirements of virtual companion applications, enabling more generalized model training.
In later self-evolution rounds, preferred online interaction samples can also be incorporated as in-context examples to further guide the LLM toward generating dialogues that better match real user preferences.

Once the dialogue text is generated, we convert the agent response into speech using a voice cloning or text-to-speech system, and then synthesize the corresponding co-speech motion sequence conditioned on the generated response speech. In this way, the pipeline produces paired multimodal interaction data consisting of dialogue context, response speech, and co-speech body motion, which can be used to train the online gesture generator.

\subparagraph{\textbf{Cold-start motion capture dataset.}}
While synthetic interaction data improves diversity and scale, training solely on synthesized motion may introduce artifacts or distribution bias. To provide high-quality grounding for the data synthesis pipeline, we additionally construct a cold-start motion capture dataset for interactive scenarios. Specifically, we invite professional actors to perform user-agent interactions according to pre-designed dialogue scripts, where the agent responses are generated or selected in advance and then enacted together with corresponding speech and co-speech gestures. This process yields paired response speech and high-quality motion sequences captured from real performers.

The resulting dataset, denoted as JIYI, serves two purposes. First, it provides a reliable supervision source for training the online gesture generator in the early stage, before sufficient online interaction data is accumulated. Second, it serves as a quality anchor for the synthetic data pipeline, helping reduce the domain gap between synthesized motion and realistic human gestures. In practice, we combine this cold-start dataset with the synthesized interaction data to train the online model.

\subparagraph{\textbf{Offline co-speech gesture generation for data synthesis.}}
To generate motion data at scale within the offline pipeline, we employ an offline co-speech gesture generator with a stronger generation capacity and relaxed latency constraints. Unlike the online gesture generator used at deployment time, the offline generator is allowed to access the complete response speech segment and can therefore synthesize more expressive and refined motion trajectories. This design is suitable for data construction, where generation quality is prioritized over inference speed.

Concretely, we first generate the agent dialogue and response speech, and then use our offline co-speech gesture generator to synthesize the corresponding co-speech motion sequence. Since the generated motions are used as training data rather than directly deployed for online interaction, we further remove low-quality samples using quality filtering and refinement. The resulting synthetic motion data provides diverse supervision for training the online gesture generator in virtual companion scenarios.

\subparagraph{\textbf{User-feedback-driven self-evolution.}}
A key advantage of our framework is that it forms a closed loop between online deployment and offline training. After deployment, the system continuously collects online interaction data, including user inputs, agent responses, generated gestures, and user feedback signals. These feedback signals may reflect user preference regarding response naturalness, synchronization, appropriateness, or overall interaction quality.

We use the collected online data to update the synthesis pipeline in a user-preference-aware manner. Specifically, preferred or high-quality interaction samples are added back to the offline data pool, i.e., \emph{Data Library} in Figure~\ref{fig:pipeline1}, while low-quality samples can be filtered out by a validation tool that can evaluate the speech-motion synchronization.
The updated Data Library is then used to regenerate (providing in-context examples for LLM models) offline dialogue synthesis in the next round or augment the training set and retrain the online gesture generator, thereby forming a self-evolving cycle. Through this iterative process, the model gradually adapts to user preferences and interaction patterns encountered during real deployment.
In practice, the preferred online interaction samples serve not only as additional supervision, but also as reference for constructing subsequent offline synthesized interactions, allowing training data distribution to gradually shift to real user preferences.

Overall, the proposed offline interactive data synthesis pipeline provides both \emph{data diversity} and \emph{continual adaptation} for online co-speech gesture generation. It not only alleviates the scarcity of virtual-companion-oriented training data, but also closes the gap between offline data construction and online deployment through user feedback.

\begin{table}[h!]
  \caption[caption]{\textbf{Comparison with state-of-the-art} methods under the strict \textit{online} setting on the BEATv2 benchmark. We highlight the best result in \textbf{Bold} and the second best in \underline{underline}.}
  \label{tab:beat}
  \begin{center}
  \resizebox{0.99\linewidth}{!}{
  \begin{tabular}{l c c c c}
  \toprule
  Method & FGD$\downarrow$ & BC $\uparrow$ & Diversity $\uparrow$ & Latency(ms)$\downarrow$ \\
  \midrule
  Semantic Gesticulator~\citep{zhang2024semantic} & $16.105$ & $ 7.246$ & $8.149$ & $+\infty$ \\
  TalkSHOW~\citep{yi2023generating} & $18.761$ & $7.167$ & $7.935$ & $8.844$ \\
  SynTalker~\citep{chen2024enabling} & $16.789$ & $7.398$ & $8.868$ & $\underline{3.336}$ \\
  EMAGE~\citep{liu2024emage} & $15.908$ & $7.035$ & $8.932$ & $5.441$ \\
  LOM~\citep{chen2025language} & $\underline{15.681}$ & $\underline{7.446}$ & $9.368$ & $22.902$ \\
  \midrule
  \textbf{\methodname}~(\textbf{base}) & $\mathbf{10.519}$ & $\mathbf{8.389}$ & $\mathbf{10.505}$ & $\mathbf{1.623}$ \\
  \bottomrule
  \end{tabular}}
  \end{center}
\end{table}

\begin{table}[t!]
  \caption[caption]{\textbf{Comparison with state-of-the-art} methods and ablation study under the strict \textit{online} setting on the JIYI dataset. We highlight the best result in \textbf{Bold} and the second best in \underline{underline}.}
  \label{tab:JIYI}
  \begin{center}
  \resizebox{0.99\linewidth}{!}{
  \begin{tabular}{l c c c c}
  \toprule
  Method & FGD$\downarrow$ & BC $\uparrow$ & Diversity $\uparrow$ & Latency(ms)$\downarrow$ \\
  \midrule
  Semantic Gesticulator~\citep{zhang2024semantic} & $2.974$ & $ 7.258$ & $9.629$ & $+\infty$ \\
  TalkSHOW~\citep{yi2023generating} & $3.703$ & $6.896$ & $9.521$ & $8.844$ \\
  SynTalker~\citep{chen2024enabling} & $3.554$ & $6.825$ & $10.996$ & $3.333$ \\
  EMAGE~\citep{liu2024emage} & $3.425$ & $6.048$ & $11.021$ & $5.441$ \\
  LOM~\citep{chen2025language} & $3.240$ & $6.991$ & $11.130$ & $22.902$ \\
  \midrule
  \textbf{\methodname}~(\textbf{base}) & $2.795$ & $7.354$ & $11.520$ & $\underline{1.623}$ \\
  \methodname~(base)~w/o audio attn. mask & $3.037$ & $ 7.027$ & $\underline{11.779}$ & $1.668$ \\
  \methodname~(base)~w/o cross-attention & $3.745$ & $6.844$ & $8.749$ & $\mathbf{1.053}$ \\
  \methodname~+ self-evolution~(1 round) & $\underline{2.419}$ & $\underline{7.395}$ & $11.532$ & $1.623$ \\
  \methodname~+ self-evolution~(2 rounds) & $\mathbf{2.231}$ & $\mathbf{7.738}$ & $\mathbf{12.509}$ & $1.623$ \\
  \bottomrule
  \end{tabular}}
  \end{center}
\end{table}

\section{Experiments}
We evaluate the proposed framework under a strict \textbf{online co-speech gesture generation} setting, where the model must generate body motion causally from streaming response speech without access to future audio. At each time step, only the current and past response speech are available to the model, while future speech information is strictly masked out. This setting is designed to reflect real-time interactive deployment, where gesture generation must satisfy both low-latency and speech-synchronization requirements.

\subsection{Experiment setup}
\label{subsection: details}

\subparagraph{\textbf{Datasets.}}
We conduct experiments on two datasets: the public BEATv2 dataset~\cite{liu2024emage} and our collected JIYI dataset. BEATv2 is a widely used benchmark for co-speech gesture generation. In alignment with the baseline setting in~\cite{liu2024emage}, we selectively use the data from the same speakers. To better reflect interactive virtual companion scenarios, we additionally construct JIYI, a high-quality interaction-oriented dataset collected from professional actors performing user-agent dialogues with synchronized speech and co-speech gestures. It contains about 6 hours
multimodal interactive data including speech, transcripts, and co-speech body motions. JIYI serves as both a cold-start dataset for model training and an evaluation benchmark for interactive settings. All motions are in SMPL-X format~\citep{pavlakos2019expressive} and consist of 30 frames per second. To further enrich the dataset, we apply the augmentation through mirroring operations during training.

\subparagraph{\textbf{Evaluation Metrics.}} Following prior work~\cite{liu2024emage,chen2024enabling,chen2025language}, we adopt the following metrics to quantitatively evaluate results: 1) Frechet Gesture Distance (\textbf{FGD}) evaluates the realism of the body gestures. 2) ~Beat Correlation (\textbf{BC}) assesses speech-motion synchronization. 3) ~\textbf{Diversity} evaluates the degree of generated motion diversity, which is calculated with the $l_1$ distance between multiple body gesture clips. Since our goal is online real-time interaction, we additionally report \textbf{Latency} (ms), defined as the average inference time of the gesture generation module per generation step under batch size 1. Unless otherwise specified, latency is measured on the same hardware platform under identical evaluation conditions and a lower value indicates better real-time responsiveness. More details about datasets and metrics are provided in \textbf{Appendix}.

\subsection{Comparison to baselines}

We compare the proposed method with representative state-of-the-art co-speech gesture generation approaches under the strict online setting described above. Since most existing methods are originally developed for offline generation, there is no standard benchmark dedicated to our online setting. Therefore, following common practice for emerging evaluation protocols, we re-implement or adapt strong prior methods as online baselines.

\subparagraph{\textbf{Baselines.}}
We adopt the following representative methods: 1) Semantic Gesticulator~\citep{zhang2024semantic}, an offline co-speech gesture generation framework, which introduces a semantic gesture retrieval module to optimize the motion sequences; 2) TalkSHOW~\citep{yi2023generating}, commonly utilized in earlier generative models for upper body co-speech gesture generation; 3) SynTalker~\citep{chen2024enabling}, the current diffusion-based method for co-speech gesture generation with competitive quality; 4) EMAGE~\citep{liu2024emage}, the competitive framework for generating full-body motions from from audio and masked gestures, which adopts masked modeling with transformer architectures; 5) LOM~\citep{chen2025language}, the recent state-of-the-art method for co-speech motion generation in the offline setting, which uses pre-trained Flan-T5-Base model~\cite{raffel2020exploring} with an encoder–decoder transformer structure. Implementation details of baseline methods are provided in \textbf{Appendix}.

\subparagraph{\textbf{Main results.}}
Tables~\ref{tab:beat} and~\ref{tab:JIYI} report quantitative comparisons on BEATv2 and JIYI, respectively. The proposed method consistently achieves the best overall balance between motion quality and real-time responsiveness. On BEATv2, our method significantly improves FGD and Diversity while achieving substantially lower latency than all compared baselines. On JIYI, which is more challenging and closer to real interactive virtual companion scenarios, our method remains competitive in motion quality and synchronization while maintaining the lowest latency. These results verify that our framework achieves a substantially better latency-quality trade-off than existing baselines when deployed in real-time interaction.

\subsection{Ablation Study}
\label{sec:ablation} 
We focus on two aspects of our framework : (1) the effectiveness of the proposed causal online gesture generation design, and (2) the benefit of the user-feedback-driven self-evolving loop.

\subparagraph{\textbf{Effect of causal audio-conditioned modeling.}}
To evaluate the role of causal speech-motion alignment, we first remove the causal audio attention mask from the cross-attention module. This variant allows the model to attend to speech features without the proposed causal masking strategy, which weakens the consistency between training and online deployment. As shown in Table~\ref{tab:JIYI}, removing the causal audio attention mask degrades both gesture quality and synchronization, leading to worse FGD and BC scores. This result indicates that strict causal masking is important for learning online-compatible speech-motion alignment, rather than being merely an inference-time constraint.

\subparagraph{\textbf{Effect of cross-modal interaction.}}
We further remove the audio-conditioned cross-attention module and replace it with a weaker interaction mechanism (prefix-conditioning, i.e., concatenating speech before motion tokens and using a causal mask to prevent future speech access) between speech and motion streams. This variant significantly degrades performance. The result verifies that explicit cross-modal interaction is essential for online co-speech gesture generation, as the model must continuously align motion prediction with the currently available response speech under incomplete future context.

\subparagraph{\textbf{Effect of self-evolution.}}
To validate the proposed self-evolving training loop, we further evaluate the model after one and two rounds of self-evolution. The base model is trained on the initial dataset. 
In each round, we progressively expand the training set with both user-preference-aware online interaction data and newly synthesized offline data. 
As shown in Table~\ref{tab:JIYI}, both one-round and two-round self-evolution improve the model over the base version, validating the effectiveness of the proposed user-feedback-driven self-evolving loop mechanism. More details and \textbf{extra data-source ablation} are provided in \textbf{Appendix}.

\begin{figure}[h]
  \centering
  \includegraphics[width=1.0\linewidth]{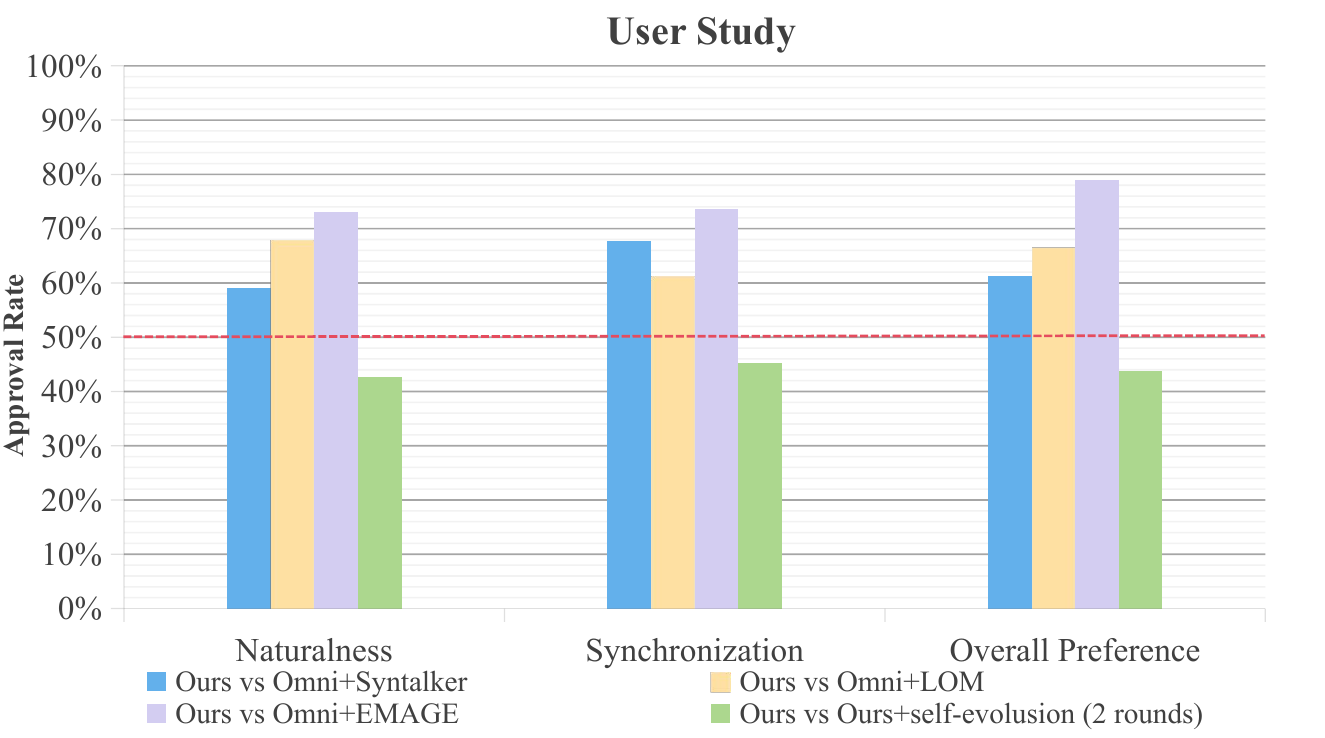}
  \caption{User Study under the pairwise preference protocol.}
  \label{fig:user_study}
\end{figure}

\begin{figure*}[h]
  \centering
  \includegraphics[width=0.9\linewidth,height=12.5cm]{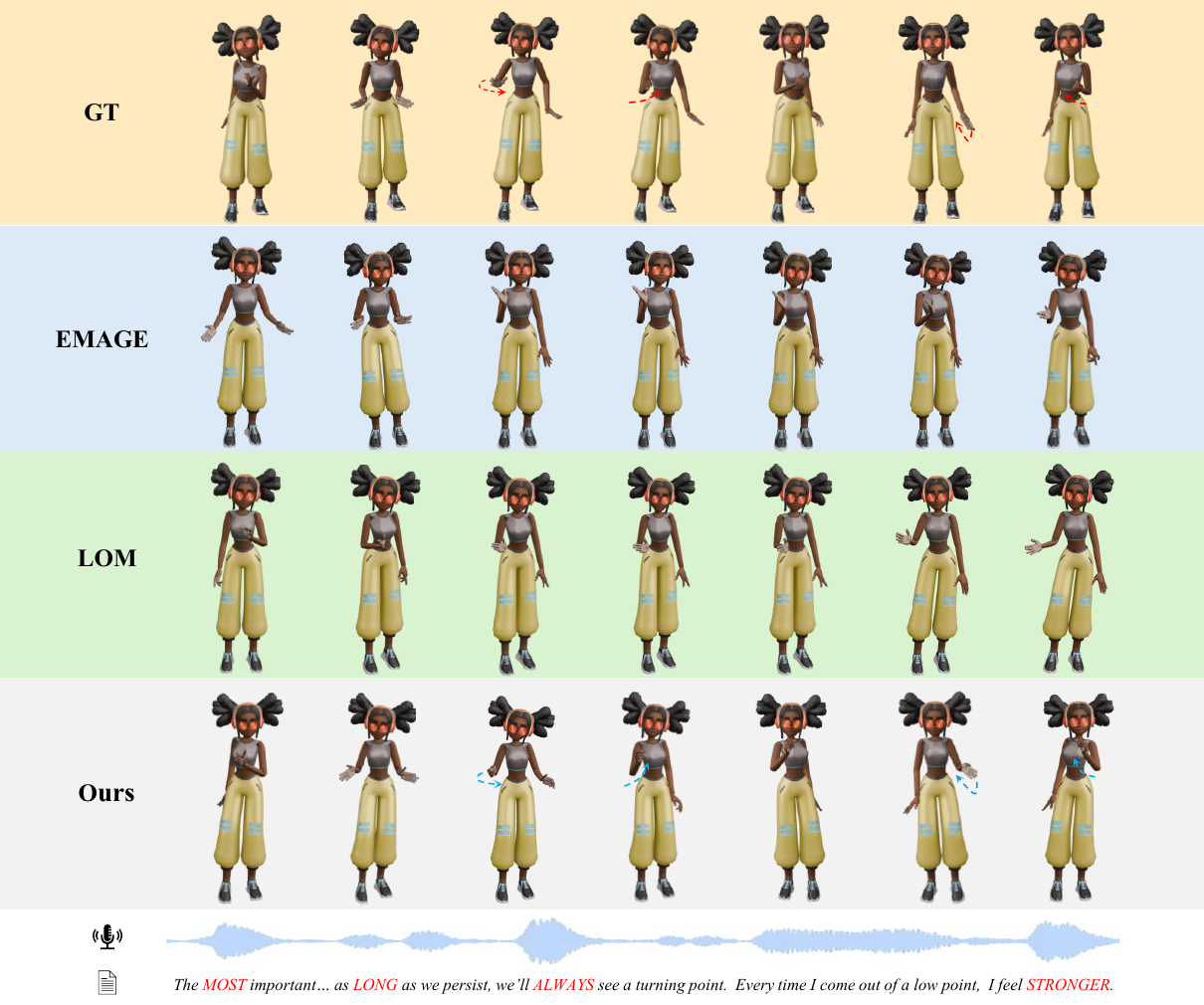}
  \caption{Qualitative comparison of online co-speech gesture generation. Given the same streaming response speech, our method reacts more promptly and generates more natural and synchronized gestures than existing baselines.}
  \label{fig:main}
\end{figure*}

\subsection{User Study}

\subparagraph{\textbf{Study setup.}}
We randomly sample interaction clips from the test set and compare the proposed method with representative strong baselines under the same strict online protocol. For each sample, all methods are conditioned on the same streaming response speech, and the generated gesture videos are rendered using the same avatar setup for fair comparison. We recruit 20 human evaluators to assess the generated results in a blind manner under the pairwise preference protocol.

\subparagraph{\textbf{Evaluation criteria.}}
Following common practice in interactive animation and gesture generation evaluation, participants are asked to rate each video from three aspects:
(1) \textbf{Naturalness}: whether the body motions appear realistic and smooth;
(2) \textbf{Speech-motion synchronization}: whether the gestures are temporally aligned with the response speech;
(3) \textbf{Overall preference}: which result is preferred in the context of real-time interaction.
If needed, we additionally ask participants to consider \textbf{responsiveness}, i.e., whether the digital human appears to react promptly and appropriately to the ongoing speech.

\subparagraph{\textbf{Results.}}
Figure~\ref{fig:pipeline1} shows that our method is consistently preferred over the compared baselines across all evaluation aspects. In particular, participants rate our method higher in motion naturalness and speech-motion synchronization, which is consistent with the quantitative improvements in FGD and BC. More importantly, our method obtains the highest overall preference in the online setting, indicating that the proposed framework produces gestures that are more suitable for real-time interactive digital humans. 

\subsection{Qualitative Evaluation and Discussion}

\subparagraph{\textbf{Qualitative Results.}} We visualize representative gesture generation results on JIYI for qualitatively comparison with strong baselines under the strict online setting. As shown in Figure~\ref{fig:main}, the proposed method generates more natural and expressive co-speech gestures with clearer temporal alignment to the response speech. Compared with existing baselines, our results exhibit smoother motion transitions, more stable body coordination, and richer upper-body and hand movements.
We further observe the proposed method better preserves gesture continuity during streaming generation, while some baselines show either over-smoothed motions or unstable local movements when future speech context is unavailable.
These qualitative differences validate the effectiveness of our method.

\subparagraph{\textbf{Limitations.}}
Although the proposed framework achieves promising results, we identify considerable areas where our model can be further improved, such as long-horizon planning under strict online constraints, generalization without seed data, fine-grained preference modeling and generalized embodiment scope. More detailed analysis and future work are provided in \textbf{Appendix}.

\section{Conclusion}

In this paper, we studied \textbf{online co-speech gesture generation} for interactive digital humans, a practically important yet underexplored setting where gestures must be generated causally from streaming response speech under strict real-time latency constraints. To address this problem, we proposed a unified real-time interactive framework that couples a streaming speech response module with a causal multimodal autoregressive gesture generator, enabling low-latency and speech-synchronous body motion generation without future speech access. 
To further support this setting, we introduced a closed-loop training pipeline tailored to virtual companion scenarios, including an offline interactive data synthesis pipeline and a user-feedback-driven self-evolution loop. This design alleviates the scarcity of virtual-companion-oriented co-speech data and supports continual adaptation to user preferences after deployment.
Extensive experiments on both public BEATv2 and our JIYI datasets demonstrate that the proposed framework achieves a better latency-quality trade-off, stronger speech-motion synchronization, and higher user preference than competitive existing baselines.


\clearpage
\bibliographystyle{ACM-Reference-Format}
\bibliography{mm}

\appendix
\clearpage

\begin{center}
{\linespread{1.5} \selectfont \textbf{\huge Supplementary Materials} \par}
\end{center}
\vspace{1em}

\section{Details of Our JIYI Dataset}
\subsection{Motion capture}

\begin{figure}[h]
  \centering
  \includegraphics[width=1.0\linewidth]{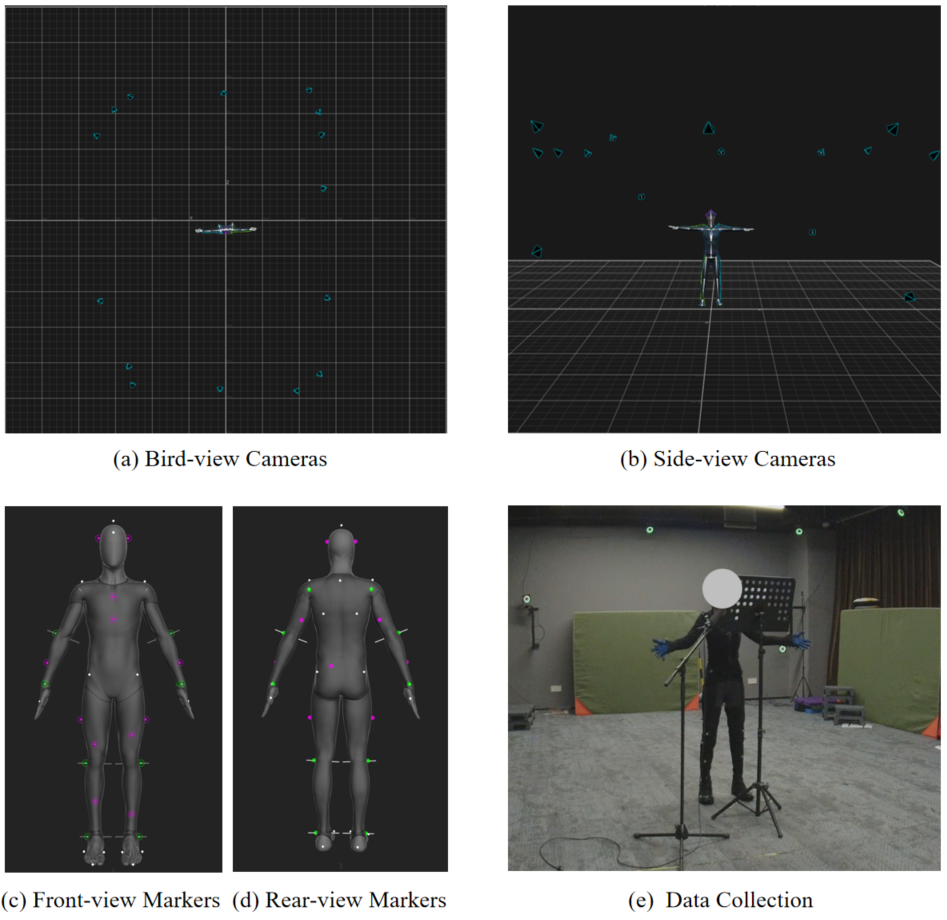}
  \caption{Cameras and Markers. We have 16 cameras around the persons: (a) cameras from bird-view and (b) cameras from side-view. We put 50 markers on one person as in (c) front-view and (d) rear-view of markers.}
  \label{fig:capture}
\end{figure}

We collect a high-quality dataset, JIYI, using the OptiTrack MoCap system equipped with 16 cameras. We invited professional actors to perform user-agent dialogues with synchronized speech and co-speech gestures. 
Our MoCap system is shown in Figure~\ref{fig:capture}. We use 16 high-resolution cameras with 120 FPS around the actor during our capture process. Figure~\ref{fig:capture} (a) and (b) show the bird-view and the side-view of
the camera positions. We apply 50 markers on one person as shown in Figure~\ref{fig:capture} (c) and (d). Figure~\ref{fig:capture} (e) illustrates the motion capture setup and environment. The actor wears motion capture suits with markers. 

\subsection{Dataset statistics}
In total, JIYI consists of about 6 hours of motion data captured at 120 FPS and we downsample the original data to 30 FPS for training. We recorded 1570 sequences in total and the train/validation/test split is 1256/157/157. Topics include weather, travel, interpersonal relationships, sports and health, games, food, professions, daily phrases and long sentences, etc. Emotions include happiness, anger, fear, sadness, calmness and rationality, etc.

\section{Experimental Details}
\subsection{Details of response module}
We adopt Qwen-Omni-3~\cite{xu2025qwen3} streaming real-time model as the backbone. The delay of its first audio token is approximately 234 ms, which meets the real-time requirements of the system.

\subsection{Details of motion tokenization}
Following~\cite{liu2024emage,chen2025language}, we use a compositional motion representation. Specifically, we divide the body into four parts represented by 6D rotations: 9 joints for the lower body $\mathbf{g}_l\in\mathbb{R}^{T\times54}$, 13 joints for the upper body $\mathbf{g}_u\in\mathbb{R}^{T\times78}$, 30 joints for the hands $\mathbf{g}_h\in\mathbb{R}^{T\times180}$, and 1 joint together with 100 expression parameters for the face $\mathbf{g}_f\in\mathbb{R}^{T\times106}$. The motion sequence is thus represented as
$
\bx=\{\mathbf{g}_f,\mathbf{g}_h,\mathbf{g}_u,\mathbf{g}_l\}.
$
\subparagraph{\textbf{VQ-VAE for online models.}}
Based on this compositional representation, we train four separate VQ-VAEs to tokenize each body part independently. For each part, the VQ-VAE encoder $\mathcal{E}$ adopts a four-layer temporal convolutional network (TCN) to extract continuous latent motion features $\mathbf{z}^{1:T}=\mathcal{E}(\mathbf{g}^{1:T})$, which are then quantized into discrete motion tokens via
\begin{equation}
\mathbf{q}^t = Q(\mathbf{z}^t) := \mathop{\arg\min}_{\mathbf{q}^k \in Q} \|\mathbf{z}^t - \mathbf{q}^k\|^2 ,
\end{equation}
where $\mathbf{q}^t$ denotes the discrete code assigned to the latent feature $\mathbf{z}^t$ with a temporal window size
$w=1$. Collectively, the quantized motion latent space is
$
Q = \{\mathbf{q}_f, \mathbf{q}_h, \mathbf{q}_u, \mathbf{q}_l\}.
$
The decoder $\mathcal{D}$ reconstructs the motion from the quantized tokens as $\hat{\mathbf{g}}^{1:T}=\mathcal{D}(\mathbf{q}^{1:T})$.
Following~\cite{chen2025language}, we train the tokenizer using reconstruction, temporal smoothness, mesh, and commitment losses:
\begin{equation}
\begin{aligned}
\mathcal{L}_{\text{total}} = &\mathcal{L}_{\text{rec}}(\mathbf{g}, \hat{\mathbf{g}}) + \mathcal{L}_{\text{vel}}(\mathbf{g}', \hat{\mathbf{g}}') + \mathcal{L}_{\text{acc}}(\mathbf{g}'', \hat{\mathbf{g}}'') + \mathcal{L}_{\text{mrec}}(\mathbf{g}, \hat{\mathbf{g}}) \\
&+ \mathcal{L}_{\text{mvel}}(\mathbf{g}', \hat{\mathbf{g}}') + \mathcal{L}_{\text{macc}}(\mathbf{g}'', \hat{\mathbf{g}}'') + \mathcal{L}_{\text{comm}}(\mathbf{g}, \mathbf{q})~,
\end{aligned}
\end{equation}
where $\hat{\mathbf{g}}'$ and $\hat{\mathbf{g}}''$ denote the first-order and second-order temporal derivatives of the reconstructed motion $\hat{\mathbf{g}}$, respectively. 
We use Geodesic loss for the pose reconstruction loss $\mathcal{L}_{\text{rec}}$, $l_1$ loss for the
velocity/acceleration losses of pose ($\mathcal{L}_{\text{vel}}$/$\mathcal{L}_{\text{acc}}$) and mesh ($\mathcal{L}_{\text{mvel}}$/$\mathcal{L}_{\text{macc}}$), $l_2$ loss for the mesh reconstruction loss $\mathcal{L}_{\text{mrec}}$ and the codebook commitment loss $\mathcal{L}_{\text{comm}}$, just like~\cite{chen2025language}.
Vertices of the SMPLX-2020 mesh computed from the pose $\mathbf{g}$ and $\hat{\mathbf{g}}$ are used to compute mesh losses.

\begin{table*}[h!]
\centering
\caption{Data composition in different rounds of self-evolution.}
\label{tab:self_evolution_data}
\begin{tabular}{lcccc}
\toprule
Round &Training Samples & Original Data & Offline Synthesized Data & Preferred Online Data \\
\midrule
Base    &1256 & 100\% & 0\%  & 0\% \\
Round 1 &2434 & 50\%  & 47\% & 3\% \\
Round 2 &3852 & 33\%  & 62\% & 5\% \\
\bottomrule
\end{tabular}
\end{table*}

\subparagraph{\textbf{RQ-VAE for offline models.}}
Due to the lower real-time requirements in offline scenarios, in this stage, we do not adopt our causal attention mask mechanism to prevent our model from knowing future speech information. In addition, we employ a more fine-grained motion representation method residual VQ-VAE (RQ-VAE) like~\cite{ng2024audio,guo2024momask,zhang2024semantic,chen2024enabling} to enhance the model's ability to represent complex motion, especially finger motions. Compared to VQ-VAE~\cite{van2017neural,zhang2023generating}, RQ-VAE improves the capacity of the quantization module by adding multiple residual quantization layers to capture motions with different granularity. Specifically, the $\mathbf{z}^t$ obtained through the VQ-VAE encoder then enters the first quantization layer $Q_1$, each vector subsequently selects its closest code entry in the layer’s codebook $C_1 = \{\mathbf{q}_1^k\}_{k=1}^K$ to obtain the first quantization code $\mathbf{q}_1^t$. The corresponding residual is then computed as $\mathbf{r}_1^t$ = $\mathbf{z}^t - \mathbf{q}_1^t$. This residual $\mathbf{r}_1^t$ is then fed into the second quantization layer $Q_2$ to select the closest code entry from the corresponding codebook $C_2$, yielding the second quantization code $\mathbf{q}_2^t$. Accordingly, $\mathbf{q}_3^t, \mathbf{q}_4^t, \cdot, \mathbf{q}_L^t$ can be obtained in the same way. In the final stage of motion encoding, all quantization codes are aggregated to form the final representation, i.e., $\mathbf{q}^t = \sum_{l=1}^{L} \mathbf{q}_l^t$. We set $L=4$ and the codebook size $C=512$ in our experiments.

\subsection{Implementation details}
On the JIYI dataset, for our online gesture generator, we adopt 12 transfomer decoder layers for our base model.
For the non-causal offline gesture generator, we set residual quantization layer $L=4$ and the codebook size $C=512$ in our experiments. Inspired by~\cite{zhang2024semantic}, we train three fine-tuning layers to refine the initially generated motion sequences layer by layer to achieve higher quality motion generation. We adopt 12 transfomer decoder layers to generate initial motion sequences and 6 transfomer decoder layers for each finetuning layer. Following prior works~\cite{zhang2024semantic}, we use a sliding history window of 120 frames for both motion and aligned audio context and do not reset at utterance boundaries, and no KV cache is used. Therefore, the computational cost is bounded by the fixed window length, which enables stable streaming latency. 
On the BEATv2 dataset, We keep the model architecture as similar as possible to ~\citep{chen2025language} for fair comparison. We conducted experiments on a GPU with 6912 CUDA cores and 108 streaming multiprocessors with 80 GB of VRAM.

\begin{table*}[h!]
\centering
\caption{Data-source ablation. All models are evaluated on the held-out test split. The offline gesture generator is non-causal and evaluated only for pseudo-label quality.}
\label{tab:data_ablation}
\resizebox{0.9\linewidth}{!}{
\begin{tabular}{l c c c c c c c c}
\toprule
Method & Training Samples & MoCap & Offline Synth. & Preferred Sup. & Preferred ICL & FGD$\downarrow$ & BC$\uparrow$ & Div.$\uparrow$ \\
\midrule
Offline Generator & 1256 & \checkmark & -- & -- & -- & 2.512 & 7.587 & 11.713 \\
\midrule
LOM base & 1256 & \checkmark & $\times$ & $\times$ & $\times$ & 3.240 & 6.991 & 11.130 \\
~+ Round 2 Full & 3852 & \checkmark & \checkmark & \checkmark & \checkmark & 2.622 & 7.301 & 11.876 \\
\midrule
Super Star base & 1256 & \checkmark & $\times$ & $\times$ & $\times$ & 2.795 & 7.354 & 11.520 \\
~+ Round 2 w/o preferred online & 3852 & \checkmark & \checkmark & $\times$ & $\times$ & 2.434 & 7.422 & 12.324 \\
~+ Round 2 w/o offline synth. & 1448 & \checkmark & $\times$ & \checkmark & -- & 2.704 & 7.390 & 11.524 \\
~+ Round 2 Full & 3852 & \checkmark & \checkmark & \checkmark & \checkmark & $\mathbf{2.231}$ & $\mathbf{7.738}$ & $\mathbf{12.509}$ \\
\bottomrule
\end{tabular}
}
\end{table*}

\subsection{More details of self-evolution pipeline}
\label{sec:supp_self_evolution}
This section provides additional details on the user-feedback-driven self-evolution pipeline.
Tables~\ref{tab:self_evolution_data} shows data composition in different rounds of self-evolution.

\subparagraph{\textbf{Overview.}}
The self-evolution pipeline is designed to progressively adapt the online gesture generation model to real deployment scenarios. Starting from an initial high-quality training dataset, we iteratively incorporate user-preference-aware online interaction samples and newly synthesized offline interaction data to expand the training set. The key idea is that a small amount of high-value online data can serve not only as additional supervision, but also as guidance for constructing a larger amount of scenario-aligned offline data.

\subparagraph{\textbf{Initial dataset.}}
Our base model is trained on an initial dataset containing 1256 sequence training samples. This dataset serves as a stable quality anchor throughout the self-evolution process and is retained in all subsequent rounds to reduce distribution drift and preserve motion realism.

\subparagraph{\textbf{Round 1 self-evolution.}}
After the first deployment round, we collect online interaction samples together with user feedback signals. Preferred online samples are selected according to user-side evaluation. These preferred samples are used in two ways:
(1) they are directly added back into the training set as user-preference-aware supervision;
(2) they are used as in-context examples in the prompt for LLM-based dialogue synthesis, guiding the generation of new human-agent dialogues that better match real interaction preferences.

The newly synthesized dialogues are subsequently converted into response speech and paired co-speech motion data through the offline synthesis pipeline. After the first round, the training set is expanded to 2434 samples, including 50\% original data, 47\% synthesized offline data, and 3\% preferred online interaction data.

\subparagraph{\textbf{Round 2 self-evolution.}}
In the second round, the updated model is deployed again and a new batch of preferred online interaction samples is collected. As in the first round, these preferred online samples are reused both as direct supervision and as in-context examples for the next round of offline dialogue synthesis. This further improves the alignment of synthesized data with user-preferred topics, interaction styles, and affective tendencies in virtual companion scenarios.

After the second round, the training set is expanded to 3852 samples, consisting of 33\% original data, 62\% synthesized offline data, and 5\% preferred online interaction data. Compared with the first round, the proportion of evolved data increases, allowing the model to progressively adapt to the target interaction domain while still preserving the original captured data as a quality anchor.

\subparagraph{\textbf{Role of preferred online samples.}}
Although the preferred online interaction data accounts for only a small fraction of the full training set, it carries high-value signals from real deployment. In our pipeline, these samples play two complementary roles. First, they provide direct supervision that reflects user preference more faithfully than generic offline data. Second, by serving as in-context examples, they influence the LLM to generate new dialogues that better align with user-preferred interaction patterns. This design turns a small amount of online preference data into a larger amount of scenario-aligned synthesized data, making the self-evolution process both data-efficient and deployment-aware.

\subparagraph{\textbf{Design rationale.}}
The mixture ratio in each round is chosen to balance three factors: 
(1) the original dataset provides stable high-quality supervision; 
(2) offline synthesized data increases topic and emotion coverage for virtual companion scenarios; and 
(3) preferred online data introduces deployment-specific preference signals. 
By keeping the original dataset in every round and gradually increasing the proportion of evolved data, the model can improve adaptation without excessively drifting away from realistic motion distributions.

\subparagraph{\textbf{Summary.}}
Overall, the self-evolution pipeline can be viewed as a preference-guided data expansion mechanism. A small amount of preferred online interaction data is leveraged both as direct supervision and as in-context guidance for LLM-based dialogue synthesis, enabling the construction of increasingly user-aligned offline training data over multiple rounds.

\subsection{Validation tool and user feedback}
\label{sec:validation}
In our experiments, user satisfaction/dissatisfaction choice is used as a lightweight preference signal to select preferred online interaction samples, rather than as a frame-level reward. These samples are used in two ways: direct preference-aware supervision and in-context examples for subsequent LLM dialogue synthesis. The validation tool filters synthesized data using multiple automatic checks, including Beat Consistency, joint-limit check, foot-skating detection, and abnormal motion/velocity thresholds. BC filters poor speech-motion rhythmic alignment, while joint-limit and foot-skating checks reject physically implausible motions. 

\subsection{Extra data-source ablation}
\label{sec:supp_data_ablation}
To isolate whether the gains come from the model or from richer data, we added extra data-source ablations on JIYI in Table~\ref{tab:data_ablation}. For "w/o preferred online", we remove preferred online interaction samples from both direct supervision and LLM in-context guidance (\textbf{ICL}), regenerate non-preference-guided synthetic samples, and match the Round-2 training size. "w/o offline synth." tests if sparse preferred online samples alone suffice without synthetic amplification. We also retrain LOM with the same Round-2 data under the same online protocol.

Results show offline synthesized data improves coverage/diversity, while preferred online feedback provides deployment-specific alignment.
Sparse preferred samples alone yield only modest gains, confirming the need for synthetic amplification. LOM benefits from Round-2 data, but remains inferior to ours, proving improvements are not solely from data scale. The non-causal offline gesture generator achieves FGD/BC/Div. of 2.512/7.587/11.713 on the held-out test split, verifying pseudo-label quality.

\begin{table}[h!]
\centering
\caption{Comparison of recent interactive systems.}
\label{tab:comparison}
\resizebox{0.99\linewidth}{!}{
\begin{tabular}{l c c c c c c }
\toprule
Method & Interactive & Online & Co-speech gesture & Streaming & Real-time & Self-evolution \\
\midrule
SOLAMI~\citep{jiang2025solami}  & \checkmark & \checkmark & $\times$ & \checkmark & \checkmark & $\times$\\
ViBES~\citep{zhang2025vibes}  & \checkmark & \checkmark & \checkmark & \checkmark & $\times$ & $\times$\\
Mio~\citep{cai2025towards}  & \checkmark & -- & $\times$ & \checkmark & --  & \checkmark\\
ProAct~\citep{zhang2026proact}  & \checkmark & $\times$ & \checkmark & \checkmark & \checkmark & $\times$\\
Super Star & \checkmark & \checkmark & \checkmark & \checkmark & \checkmark & \checkmark\\
\bottomrule
\end{tabular}
}
\end{table}

\subsection{Details of baselines}
This section provides additional implementation details on how we implement the existing baseline methods.

\subparagraph{\textbf{Baseline adaptation on BEATv2.}}
On the BEATv2 dataset, we use their official code. During inference, we mask unknown future audio information so that each model only has access to the current and past response speech, following our strict online protocol. This setting enables us to evaluate the practical online deployment capability of existing methods without modifying their original training procedure.

\subparagraph{\textbf{Baseline adaptation on JIYI.}}
On the JIYI dataset, we further construct causal versions of the baseline methods for fairer comparison under the online setting. Specifically, we apply causal masking to their speech conditioning pathway so that the models only observe current and past response speech during training and inference. For methods originally based on sequence-to-sequence or transformer-style generation, this corresponds to replacing full-context conditioning with causal conditioning under the same online protocol. We keep the remaining model structures unchanged as much as possible to preserve the original design of each baseline. This difference is mainly due to practical availability: on BEATv2 we follow the original code for reproducibility, while on JIYI we retrain causal versions to better evaluate baseline performance under the online setting on our interaction-oriented dataset.

\subparagraph{\textbf{Training protocol.}}
We follow the same training schedule and train each model for 200 epochs. Unless otherwise specified, the input and evaluation protocols are kept consistent across methods to ensure fair comparison under the strict online setting.

\subparagraph{\textbf{Remarks on fairness.}}
We note that the compared baselines were not originally proposed for online co-speech gesture generation. Therefore, our goal is not to reproduce their best offline performance, but to evaluate how well they can be adapted to a practically relevant online setting where future speech is unavailable. Using masked inference on BEATv2 and causal retraining on JIYI allows us to compare all methods under a unified online protocol.

\subparagraph{\textbf{Relation to recent real-time or streaming gesture generation.}}
A few recent studies~\cite{ji2026discoforcing, krome2023towards, abel2024towards} have started to explore online applicability, low-latency gesture synthesis~\cite{deng2026u,zhan2026umo}, or streaming motion generation~\cite{xiao2025motionstreamer}. However, to the best of our knowledge, there is still no standard baseline in the last two years that is strictly aligned with our task setting: full-body 3D online co-speech gesture generation for interactive digital humans, conditioned on streaming response speech, under strict causal and real-time constraints.
Recent works differ from our setting in at least one of the following aspects: 
(1) they are not designed for speech-conditioned \emph{co-speech} gesture generation;
(2) they do not model \emph{streaming agent response speech} in interactive digital humans;
(3) they do not follow a strict online protocol where future speech is completely unavailable; or 
(4) they target broader reaction synthesis or general streaming motion generation rather than speech-synchronous gesture generation itself.
Therefore, in our experiments, we mainly compare with representative state-of-the-art co-speech gesture generation methods and adapt them under the same strict online protocol. We believe this comparison is currently fair and most relevant way to evaluate progress on the core task studied in this paper. We also compare recent interaction systems that are related but not directly comparable in Table~\ref{tab:comparison}.

\subsection{Details of Metrics}

\subparagraph{\textbf{Fréchet Gesture Distance (FGD).}}
FGD~\cite{yoon2020speech} measures the degree of closeness between the distribution of generated body gestures and that of the ground-truth gestures. Similar to the perceptual metric commonly adopted in image generation, FGD is computed based on latent representations extracted by a pretrained network:

\begin{equation}
\mathrm{FGD}(g, \hat{g}) = \|\mu_r - \mu_g\|^2 + \mathrm{Tr}\left(\Sigma_r + \Sigma_g - 2(\Sigma_r \Sigma_g)^{1/2}\right),
\tag{11}
\end{equation}
where $\mu_r$ and $\Sigma_r$ denote the mean and covariance of the latent feature distribution $z_r$ extracted from real gestures $g$, while $\mu_g$ and $\Sigma_g$ denote the mean and covariance of the latent feature distribution $z_g$ extracted from generated gestures $\hat{g}$. Following~\cite{liu2024emage}, we adopt an autoencoder pretrained on gesture data, which consists of a Skeleton CNN (SKCNN) based encoder and a Full CNN-based decoder.

\subparagraph{\textbf{L1 Diversity.}}
A higher L1 Diversity~\cite{li2021audio2gestures} indicates greater variation among the generated gesture clips. We compute the average L1 distance over $N$ motion clips as follows:

\begin{equation}
\mathrm{L1\ div.} = \frac{1}{2N(N-1)} \sum_{t=1}^{N} \sum_{j=1}^{N} \left\| p_t^j - \hat{p}_t^j \right\|_1,
\tag{12}
\end{equation}
where $p_t$ denotes the joint positions at frame $t$. Diversity is evaluated over the entire test set. 

\subparagraph{\textbf{Beat Constancy (BC).}}
A higher BC indicates better rhythmic alignment between the generated gestures and the audio beats. The beginning of speech is identified as the audio beat and the local minima of the velocity of the upper body joints (excluding fingers) is considered as the motion beat. The synchronization between audio and gesture beats is computed as

\begin{equation}
\mathrm{BC} = \frac{1}{|g|} \sum_{b_g \in g} \exp\left(- \frac{\min_{b_a \in a} \|b_g - b_a\|^2}{2\sigma^2}\right),
\tag{13}
\end{equation}
where $g$ and $a$ denote the sets of gesture beats and audio beats, respectively.

\subsection{Details of User Study}
\label{sec:supp_user_study}

This section provides additional details of the user study protocol, including participant setup, evaluation criteria, and result collection.

\subparagraph{\textbf{Study goal.}}
The user study is designed to evaluate the perceptual quality of generated co-speech gestures in real-time interaction scenarios. While objective metrics such as FGD, BC, Diversity, and Latency quantify motion realism, synchronization, and efficiency, they cannot fully reflect human perception of gesture naturalness and interaction quality. Therefore, we additionally conduct a subjective evaluation to assess whether the generated gestures are visually natural, speech-synchronous, and preferable in practical online interaction.

\subparagraph{\textbf{Evaluation setting.}}
We randomly sample a subset of test interaction clips from the evaluation set and render the generated gestures of different methods using the same avatar, camera view, and visualization setup. All methods are evaluated under the same strict online protocol, i.e., only the current and past response speech are available during generation. To ensure fairness, each compared method is conditioned on the same streaming response speech for each test sample.

\subparagraph{\textbf{Participants.}}
We recruit 20 human participants to evaluate the rendered videos in a blind manner. The identities of the compared methods are hidden, and the presentation order of videos is randomized to reduce bias. Each participant independently watches the generated clips and provides ratings according to the predefined evaluation criteria.

\subparagraph{\textbf{Evaluation criteria.}}
Participants are asked to evaluate each result from the following aspects:
\begin{itemize}[leftmargin=*]
    \item \textbf{Naturalness}: whether the generated gestures appear realistic, smooth, and human-like.
    \item \textbf{Speech-motion synchronization}: whether the body motions are temporally aligned with the response speech.
    \item \textbf{Responsiveness}: whether the digital human appears to react promptly in the online interaction setting.
    \item \textbf{Overall preference}: the overall quality of the generated result in the context of real-time interactive digital humans.
\end{itemize}

\subparagraph{\textbf{Pairwise preference protocol.}}
For pairwise preference evaluation, participants are shown two rendered videos side by side, generated by two different methods from the same response speech input. The left-right order is randomized. Participants are asked to select the preferred result according to one of the above evaluation aspects. The final preference ratio is computed as the proportion of times one method is preferred over another across all comparisons.

\subparagraph{\textbf{Result interpretation.}}
The subjective evaluation complements the quantitative results reported in the main paper. In particular, Naturalness is related to motion quality, Speech-motion synchronization corresponds to BC, and Responsiveness is closely associated with Latency. Higher Overall preference indicates that the generated gestures provide a better user experience in real-time interaction. In our experiments, the proposed method is consistently favored by participants, supporting the effectiveness of the proposed online gesture generation framework.

\subparagraph{\textbf{Remarks.}}
We note that subjective evaluation is particularly important for interactive digital humans, since low-latency deployment alone is insufficient if the generated gestures appear unnatural or poorly synchronized. The user study therefore serves as an essential complement to the quantitative analysis in the main paper.

\section{Prompt Template for Dialogue Synthesis}
\label{sec:supp_prompt}

This section provides additional details on the prompt template used in the offline interactive data synthesis pipeline.

\subparagraph{\textbf{Overview.}}
Our goal is to synthesize human-agent dialogues tailored to virtual companion scenarios. To improve controllability and diversity, the prompt is constructed from two sources: (1) topic and emotion cues sampled from the subject corpus, and (2) preferred online interaction samples used as in-context examples during self-evolution rounds. The topic-emotion cues provide high-level semantic and affective control, while the in-context examples help align newly generated dialogues with user-preferred interaction styles observed in deployment.

\subparagraph{\textbf{Base prompt template.}}
In the initial offline synthesis stage, each prompt is constructed from a sampled topic segment and a sampled emotion segment. A generic template is shown below:

\begin{quote}
\small
You are asked to generate a natural dialogue between a human user and a virtual companion agent.

Topic: [TOPIC]

Emotion style of the agent: [EMOTION]

Requirements:
(1) The dialogue should be natural, coherent, and suitable for daily interaction.
(2) The user should express a concern, question, feeling, or life situation related to the topic.
(3) The agent should respond in a way that matches the given emotion style.
(4) The agent should provide emotional value, such as comfort, encouragement, empathy, or companionship, rather than only factual information.
(5) Keep the conversation concise but expressive, and ensure that the agent response is suitable for co-speech gesture generation.

Please output a multi-turn dialogue in the format:
User: ...
Agent: ...
\end{quote}

We explicitly encourage the LLM to produce emotionally expressive and interactionally rich agent responses, since such responses provide more informative speech conditions for subsequent co-speech motion synthesis.

\subparagraph{\textbf{Prompt template with in-context examples.}}
During self-evolution, preferred online interaction samples are additionally inserted as in-context examples to guide the next round of dialogue synthesis. A generic template is shown below:

\begin{quote}
\small
You are asked to generate a natural dialogue between a human user and a virtual companion agent.

Below are examples of preferred interactions:
[EXAMPLE 1]

[EXAMPLE 2]

...

Now generate a new dialogue with the following settings:

Topic: [TOPIC]

Emotion style of the agent: [EMOTION]

Requirements:
(1) The dialogue should be natural, coherent, and suitable for daily interaction.
(2) The user should express a concern, question, feeling, or life situation related to the topic.
(3) The agent should respond in a way that matches the given emotion style.
(4) The agent should be supportive, emotionally appropriate, and engaging.
(5) The new dialogue should not copy the examples verbatim, but should follow a similar interaction style.
(6) The agent response should be expressive and suitable for subsequent co-speech motion generation.

Please output a multi-turn dialogue in the format:
User: ...
Agent: ...
\end{quote}

\subparagraph{\textbf{Formatting of in-context examples.}}
Each preferred online sample is formatted as a short user-agent interaction snippet and inserted into the prompt as a few-shot example. In practice, these examples reflect user-preferred topics, emotional tone, and interaction styles. This allows a small amount of preferred online interaction data to guide the construction of a larger amount of scenario-aligned synthesized dialogues in subsequent self-evolution rounds.

\subparagraph{\textbf{Design rationale.}}
The prompt design serves two purposes. First, explicit topic and emotion conditions improve controllability and diversity of synthesized dialogues. Second, in-context examples from preferred online interactions help the LLM better match real user expectations in virtual companion scenarios. As a result, the synthesized dialogues become progressively more aligned with user-preferred interaction patterns and provide better training data for the online co-speech gesture generation model.

\section{Limitation and Future Work}
\label{sec:supp_limitation}

Although the proposed framework achieves promising results in interactive digital humans, through comparative experiments, we identify considerable areas where our model can be further improved, as outlined below:

\subparagraph{\textbf{Long-horizon planning under strict online constraints.}}
Our method is designed for strict online generation, where only the current and past response speech are available. Although this design is necessary for real-time deployment, it also limits the model's ability to leverage long-range future context. As a result, when the response involves rapid topic shifts, highly dynamic prosody changes, long-form emotionally evolving utterances, or highly anticipatory gestures, the generated gestures are still less globally structured than those produced by fully offline models with complete future speech.

\subparagraph{\textbf{Dependence on synthesized interaction data.}}
To address the scarcity of virtual-companion-oriented training data, we introduce an offline interactive data synthesis pipeline. While this substantially improves data diversity and scenario coverage, synthesized dialogues and motions may still exhibit distribution gaps from real human interaction. Although we mitigate this issue by retaining the original captured dataset as a quality anchor and incorporating preferred online interaction samples through self-evolution, the domain gap between synthesized and real interactive behaviors is not fully eliminated.

\subparagraph{\textbf{Generalization without seed data.}}
We propose the offline  data synthesis pipeline to address data scarcity. However, our framework assumes a small high-quality seed dataset, or alternatively a relevant public co-speech dataset, to initialize the offline synthesis pipeline and reduce domain shift. Generalization without any domain-specific seed data is not fully explored. In addition, broader speaker-specific motion-style generalization is also worth exploring.

\subparagraph{\textbf{Lightweight preference modeling.}}
Our self-evolution mechanism uses preferred online interaction samples as both additional supervision and in-context examples for subsequent dialogue synthesis. This design is effective and practical, but it remains a relatively lightweight form of preference modeling. In particular, the current framework does not explicitly learn a dedicated user preference model, reward model, or fine-grained adaptation policy. Therefore, the system may not fully capture subtle or personalized long-term user preferences.

\subparagraph{\textbf{Generalized embodiment scope.}}
Our current framework mainly focuses on online co-speech body motion generation conditioned on streaming response speech. However, real-world digital humans typically require tighter integration of multiple embodied channels, such as facial micro-expressions, eye gaze, head attention, turn-taking behavior, and broader environmental grounding. Extending the current framework toward more holistic multimodal embodiment remains an important direction for future work. Additionally, the digital human can be replaced with humanoid robots~\cite{zhang2026proact}, laying the groundwork for real-world deployment.

\subparagraph{\textbf{Future work.}}
There are several promising directions for future research. First, it would be valuable to explore more advanced online planning mechanisms that preserve strict real-time responsiveness while improving long-horizon motion coherence. Second, richer preference-aware adaptation strategies could be introduced, such as explicit preference modeling, personalized dialogue-motion synthesis, or more fine-grained data selection and weighting during self-evolution. Third, expanding the framework to more comprehensive virtual companion settings, including richer affective interaction, multimodal grounding, and long-term user adaptation, may further improve the realism and emotional engagement of interactive digital humans. Nevertheless, we believe the proposed framework provides a practical step toward more responsive, expressive, and adaptive digital humans in real-world interaction.

\end{document}